# High Accuracy Traffic Light Controller for Increasing the Given Green Time Utilization

Maythem K. ABBAS[1,a,*], Mohd N. KARSITI[1,b], Madzlan NAPIAH[2,c], Brahim B. SAMIR[3,d], and Marwan Al-Jemeli[1,e]

[1]University Tecknologi PETRONAS, Electrical and Electronics Engineering Department, Malaysia

[2] University Tecknologi PETRONAS, Civil Engineering Department, Malaysia

[3]Mathematics and computer science department, College of science, Al-Faisal University, Saudi

[a]eng_maythem_84@yahoo.com, [b]nohka@petronas.com.my, [c]madzlan_napiah@petronas.com.my, [d]samir.brahim@gmail.com, [e]shaheen.marwan@gmail.com

*Maythem K. Abbas

H/P# 006-01-733-70965 (Malaysia)

**Abstract.** Traffic congestion has become one of the major problems in the urban cities according to the increasing number of vehicles in those cities, obsolete technologies used on the roads of those cities, inappropriate road design, and many other reasons. So, that has urged the need for a more accurate traffic light controlling system; one that will help in maintaining high stability at all levels of demand. This paper introduces a dynamic traffic light phase plan protocol for Single-Isolated Intersections. The developed controlling method was compared with four other methods and showed a good performance in terms of reducing the average and maximum queue lengths, optimizing the given green time amount as needed, and increased the intersection's throughput (increased the given green time utilization). In addition, it maintained a good traffic light stability at all levels of demand.

1. Introduction

Traffic congestion has become one of the major problems in the urban cities according to the increasing number of vehicles in those cities, obsolete technologies used on the roads of those cities, inappropriate road design, and many other reasons. So, that has urged the need for a more accurate traffic light controlling system; one that will help in maintaining high stability at all levels of demand.

Generally, a traffic light controlling system consists of two main entities as shown in Figure 1: Traffic Light Controller (TLC) and Traffic Light Display entity (TLD). The latest solutions have suggested having a third entity, Road Status Data Collector (RSDC), in addition to the classic two entities as shown in Figure 2. The job of the additional entity would be to collect real-time data about the approach's lanes and deliver them to the traffic light controller which would make a decision about the next phase plan based upon those collected data.

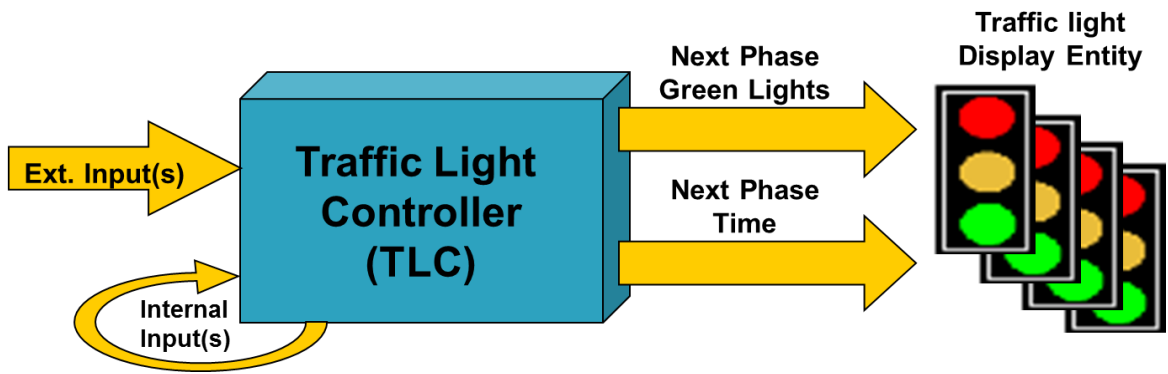

Figure 1: Basic Traffic Light System Entities [1] and [2]

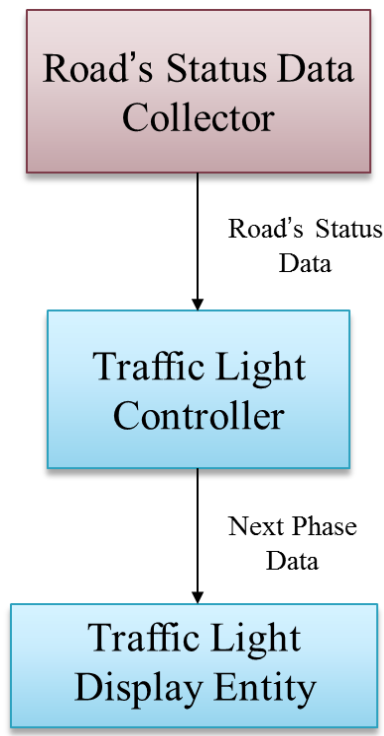

Figure 2: Nowadays Traffic Light System Entities

The aim of this work has been to increase the traffic light phase plan decision making accuracy and maintain high stability at all of the levels of demand. That urged researching the ability of finding solutions for the problems of what type of road data are needed to be collected, how they will be delivered to the traffic light controller, and how they will be used, within the TLC, to make a decision about the next phase plan. In addition, a new dynamic mechanism for the queue length detection was developed and illustrated within this paper.

**2. Literature Review**

In this section, a set of the previously proposed approaches for intelligent Traffic light Management systems will be briefly illustrated with highlighting their downsides. Centrally controlled city traffic lights was one of the very first proposed solutions to overcome the isolated pretimed traffic lights' downsides as in [3], which was very costly and slow because of using rented telephone lines for transmitting the roads' data to the central controller. Then a newer suggestion came up which estimates the vehicles average speed and builds the traffic light phase plan upon it. The results of such a system will not be accurate as it depends on estimation not detection [4].

Some other solutions were proposed to avoid or to reduce traffic jams on roads via propagating the traffic data for drivers hoping that they would change their path and avoid the congested roads [5, 6]. According to our opinion, this solution will not necessarily solve the congestion problem on the road as it might fail in some cases, for example, when many drivers want to approach a destination which has only one way to reach to; the proposed solution will not help the driver to avoid the congestion nor ease the traffic flow.

Many researchers have argued that an image processing solution would be optimum to solve the congestion problem, such as [7], [8], [9], [10], [11], [12], [13], and [14]. Using a camera to capture the road picture or a video and then analyze the captured pictures/video will not always work as it would fail during the heavy raining, foggy, or sand storm weather, and at very dark or unlined road. While some others have argued that using Global Positioning System (GPS) would be the good solution for the traffic problem, [15]. As according to our opinion, it won't be an optimum choice for urban cities with high buildings.

Finally, several studies have found Mathematical solutions for the congestion problem which have proven as better solutions than the above, such as [16], [17], [18], and [19]. Nevertheless, it has been noticed that most of those solutions face instability when a high level of demand to use the intersection occurs. Further, they face inaccuracy in decision making and that is mainly because of the incomplete list of variables that they collect and base their decisions upon or because of using the wrong control algorithm. That has motivated us to find a better solution to overcome those works' downsides. The most related and latest approaches to ours were both of [18] and [19]. In this paper, we will refer to them as NM1 (New-Method-1) and NM2 (New-Method-2), respectively.

**3. The Developed Traffic Light System**

Just like other latest solutions, the developed traffic light system consists of three main entities: Road's Status Data Collecting entity, Traffic Light Controlling entity, and the Traffic Light Display entity. The first entity would collect data about the vehicle types (civilian or special vehicles), its status (On-Duty or free) and its position (on which lane) which will lead to determining the queue lengths of the lanes. Those data will be forwarded to the traffic light controller which will make the decisions regarding the next traffic light phase plan then forward it to the TLD to be implemented there. When the time comes to switch to a new phase, the TLC will be triggered to produce a new phase plan based on a new set of data, and so on. The intersection considered in this paper was a four leg intersection as that shown in **Figure 3**.

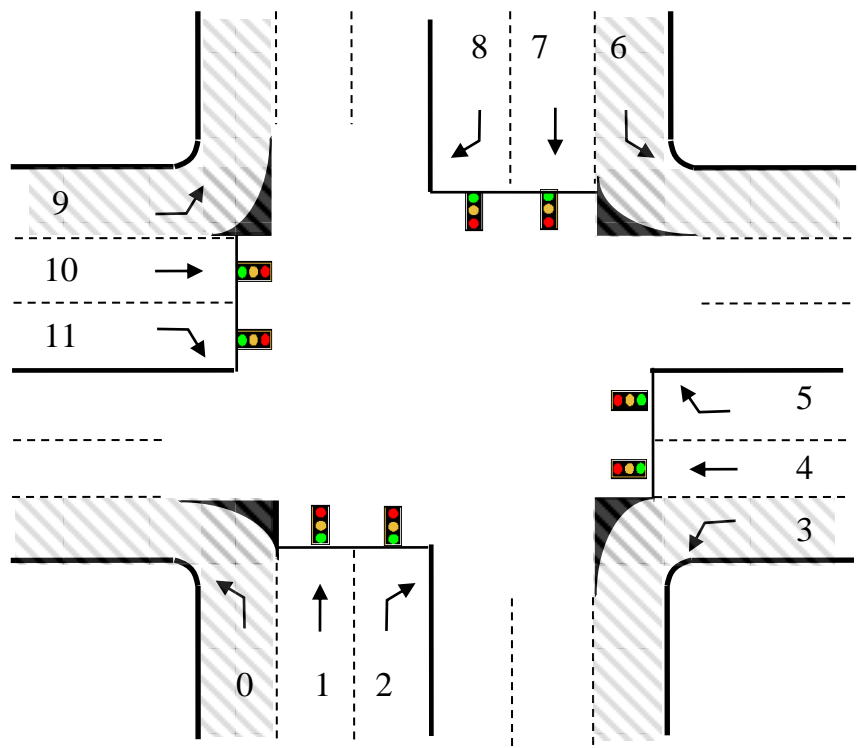

**Figure 3: Standard Four-Legs Intersection**

3.1. ROAD'S STATUS DATA COLLECTING SYSTEM

As shown in Figure 4, a set of road side equipment named as RSE were placed on the road side and connected in a serial manner. They were separated by 150 meters starting from the traffic light stopping line backwards. All the RSEs were connected to a belt of sensors except the RSE near by the traffic light stopping line was connected to two sensor belts; those were separated by 7 to 10 meters. The duty of those sensor belts depended on their position on the road. As in the setup shown in Figure 4, the belts connected to the first three RSEs (RSE-1 through RSE-3) served as queue length detectors. While, the first belt was connected to RSE-4 (on the left), which was placed just a few meters from the traffic light stopping line, and served as the first vehicle arrival detector. The other belt, which was placed just after the traffic light stopping line, served as the leaving vehicle detector.

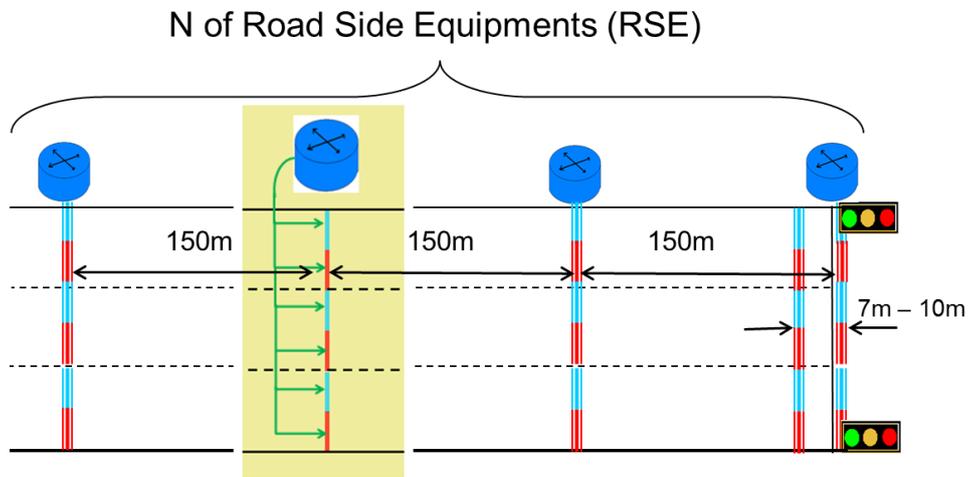

Figure 4: Road's Status Data Collecting System Setup

During the movement of a vehicle on a road, the data of that vehicle was collected through the VANET communication among the vehicle and the road side equipment (RSE). While, the position data was collected through belts placed on the road.

3.2. THE DYNAMIC QUEUE LENGTH DETECTION MECHANISM

The developed RSDC system has the ability to activate or deactivate the sensor segments of the belts according to the queue length of the lanes. In the example shown in Figure 5, initially, all of the sensor belt segments for the RSE-3 were active while those for RSE-1 and RSE-2 were inactive. As long as the queue lengths of the lanes were less than 25 vehicles, then the three segments of the sensor belt for the RSE-3 stayed active. In other words, the queue length detection for the three lanes was under the responsibility of RSE-3. While, as soon as any of the queue lengths reached the maximum number of vehicles, they were inserted within the 150 meters (averagely, 25 vehicles). As in the second lane of the down side of Figure 5, the RSE-3 would send a message to the RSE-2 to handover the responsibility of the queue length detection for the second lane to RSE-2. The responsibility of the second lane's queue length detection would be given back to RSE-3 whenever its queue became less than 25 vehicles.

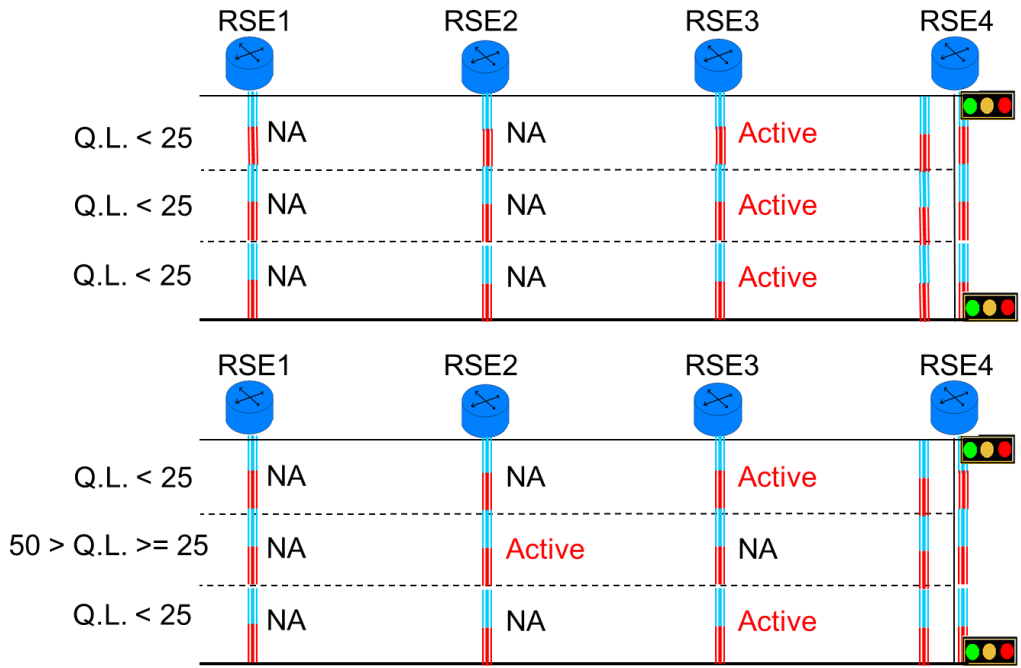

**Figure 5: The Dynamic Queue Length Detection Mechanism Functionality Example**

3.3. THE TRAFFIC LIGHT CONTROLLER (TLC)

When the time came to change the traffic light phase, the traffic light controller would receive the latest collected data for the intersection approaches and started the process of producing a new traffic light phase plan. The traffic light phase plan consisted of two main values: the index of the next phase green lights, and the phase time.

Figure 6 shows the internal architecture of the TLC which mainly had 3 blocks. The first two blocks were the lanes' load Calculation block and Graph route decision making block, and they helped in determining the first decision, the next phase green lights. While, the last block was the next phase time decision maker which performed its job based on the first decision made.

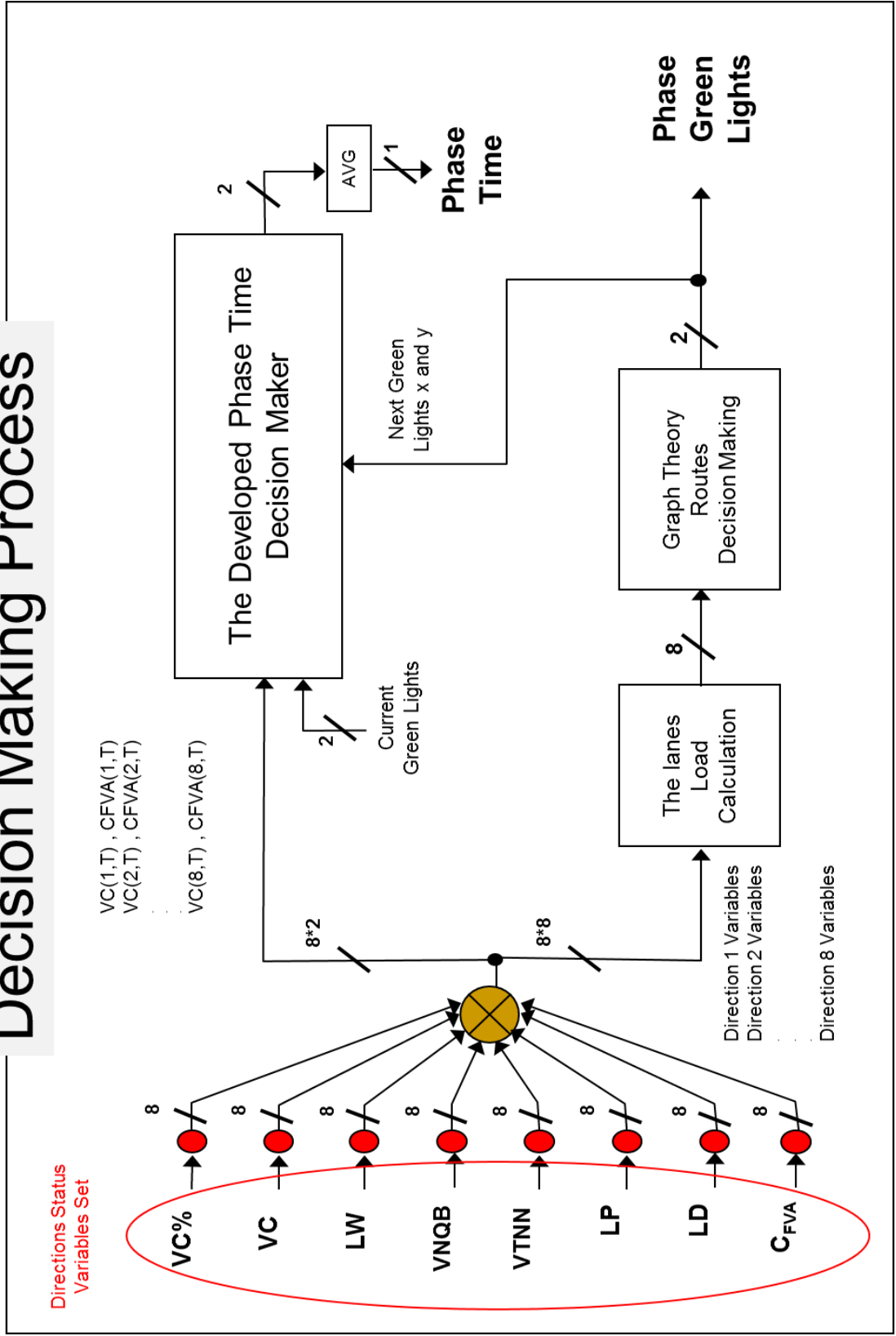

**Figure 6: The Traffic Light Controller (TLC) Internal Architecture Diagram**

### 3.3.1. Lanes' Load Calculation Block

The whole set of the intersections' collected data was forwarded to the load calculation block to be substituted in the lanes' load calculation equation, as in Figure 7.

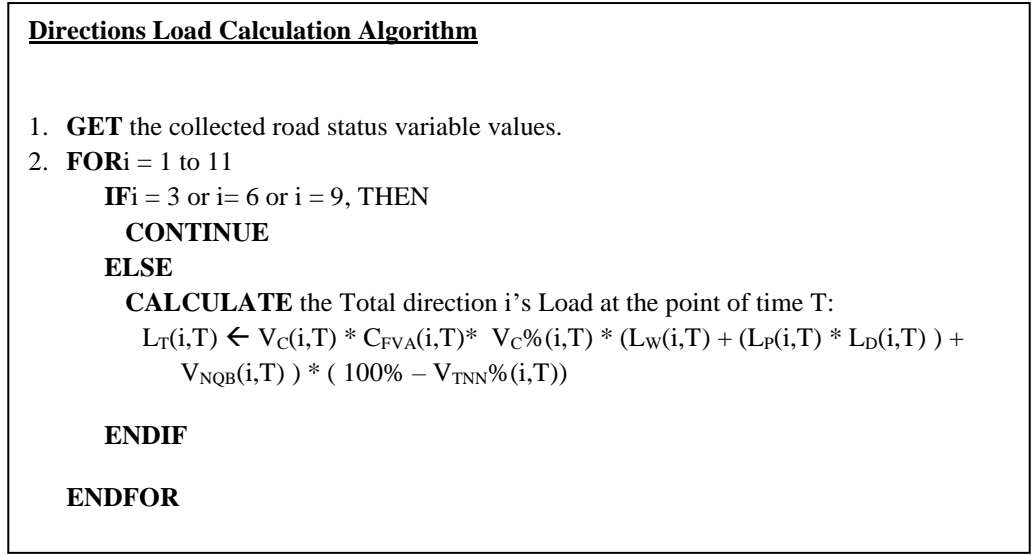

**Directions Load Calculation Algorithm**

1. **GET** the collected road status variable values.
2. **FOR** i = 1 to 11
    **IF** i = 3 or i= 6 or i = 9, THEN
      **CONTINUE**
    **ELSE**
      **CALCULATE** the Total direction i's Load at the point of time T:
      $L_T(i,T) \leftarrow V_C(i,T) * C_{FVA}(i,T) * V_C\%(i,T) * (L_W(i,T) + (L_P(i,T) * L_D(i,T)) + V_{NQB}(i,T)) * (100\% - V_{TNN}\%(i,T))$
    **ENDIF**
  **ENDFOR**

Figure 7: Lane's Load/Weightage Calculation Algorithm

The variables used to calculate each direction's load/weightage will be defined in this section. Starting with $V_C(i,T)$, which can be defined as the Vehicle Count confirmed to be within the Queuing area of direction i at the point of time T, the first vehicle arrival confirmation Flag on direction i at the point of time T has been abbreviated as $C_{FVA}(i,T)$. Whereas, $V_C\%(i,T)$ stands for what percentile of the vehicles' first queuing area of the direction i's road was occupied at the point of time T. When the first vehicle arrived to a red traffic light, a timing counter started counting the Waiting time ($L_W(i,T)$) for the first vehicle in the queue of direction i's road at the point of time T. For detecting the emergency vehicle's existence, two variables were collected; these were the vehicle's priority $L_P(i,T)$ and the flag $L_D(i,T)$ for the special vehicle (driving on direction i at the point of time T) whether it was on Duty (LD =1) or not (LD =0). Two variables were collected for integration purposes; those were the $V_{NQB}(i,T)$ (Vehicle's Total Number Queuing on the Back-road traffic lights, those leading to the direction i's road at the point of time T), and the (100% - $V_{TNN}\%(i,T)$) (how much percentile of the next road (the road which received vehicles coming from the direction i)) was instantly occupied by vehicles at the point of time T.

3.3.2. Graph Route Decision Making Block

As shown in Figure 6, the output of the load calculation block was 8 weightage values for those forwarded to the next block which was the graph route decision making block. The main functionality of this block was to choose which two lanes should be green in the next phase.

Within this block, the physical intersection was mapped into a graph named the Signalized Intersection Graph (SIG) where each lane/direction was presented as a node of the graph with a capital letter as a name. This is as shown in Figure 8. According to the intersection traffic light rules, each lane had a set of intersecting relations with 4 more lanes/directions: those were given small letters as names. These relations were dual side relations where the value of each relation depended on its direction and the weightage of the destination node. For example, in case the phase was moving from node A to node C, then the relation (a) had the value of the destination node weightage (Node C). While, if the movement was from node C to node A, then the value of relation (a) equaled the weightage of node A.

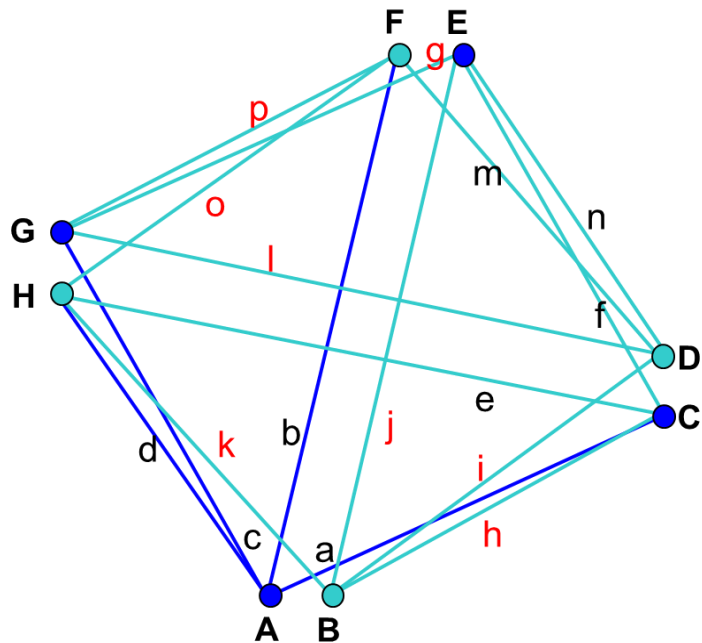

**Figure 8: Signalized Intersection Graph (SIG)**

As can be seen in Figure 9, the next phase, green light determination process, started with initializing the neighbor lists of each node. Then assigning the calculated weightage of each lane in the previous block to the responsive nodes on the SIG and accordingly calculating the relations' values. The next step was listing down the available full-Mesh element-to-element pairs between the neighbours' list members of the two currently green adjacent node lists. This represented all of the probabilities of the next phase combinations. However, there were some undesired combinations; those might be duplicated (e.g., CH and HC) or intersected (AC and BH) or paired to itself (AA or BB, etc.). This is why those combinations were not considered among the next phase probabilities and were eliminated. The final step was to choose the two nodes of the combination with the highest weightage to act as the next phase green lights.

---

**Next Green Phase Light Algorithm**

1. **INIT** Adjacent Node (Neighbor) List of:
    Node A ← C,F,G,H
    Node B ← C,D,E,H
    Node C ← A,B,E,H
    Node D ← B,E,F,G
    Node E ← B,C,D,G
    Node F ← A,D,G,H
    Node G ← A,D,E,F
    Node H ← A,B,C,F

2. **SET** A ← $L_T(1,T)$ , B ← $L_T(2,T)$, C ← $L_T(4,T)$ , D ← $L_T(5,T)$ , E ← $L_T(7,T)$, F ← $L_T(8,T)$, G ← $L_T(10,T)$, H ← $L_T(11,T)$

3. **DETERMINE** the available full-Mesh element-to-element pairs from the two currently green's adjacent node lists.

4. **ELIMINATE** the pair-to-itself combinations.

5. **ELIMINATE** the intersected (unavailable) pairs.

6. **ELIMINATE** any duplicated pairs.

7. **DECENDING SORT** the rest of the pairs in the list.

8. **SET** the first pair two elements on the list as the next phase green lights.

---

**Figure 9: The Developed Next Phase Green Light Decision Making algorithm**

3.3.3. Next Phase time decision making Block

After determining the two green traffic lights for the next phase (those indexes are x and y in Figure 6 and Figure 10), it was the time to calculate for how long they would stay green. This was why, the indexes x and y were inserted as inputs to the block of the next phase time determination to participate in its calculations.

In this stage, a portion of the full cycle time was taken off and given to lane-x and another portion was given to lane-y according to the ratios of their queue lengths to the total summation of the competing (intersecting) queue lengths.

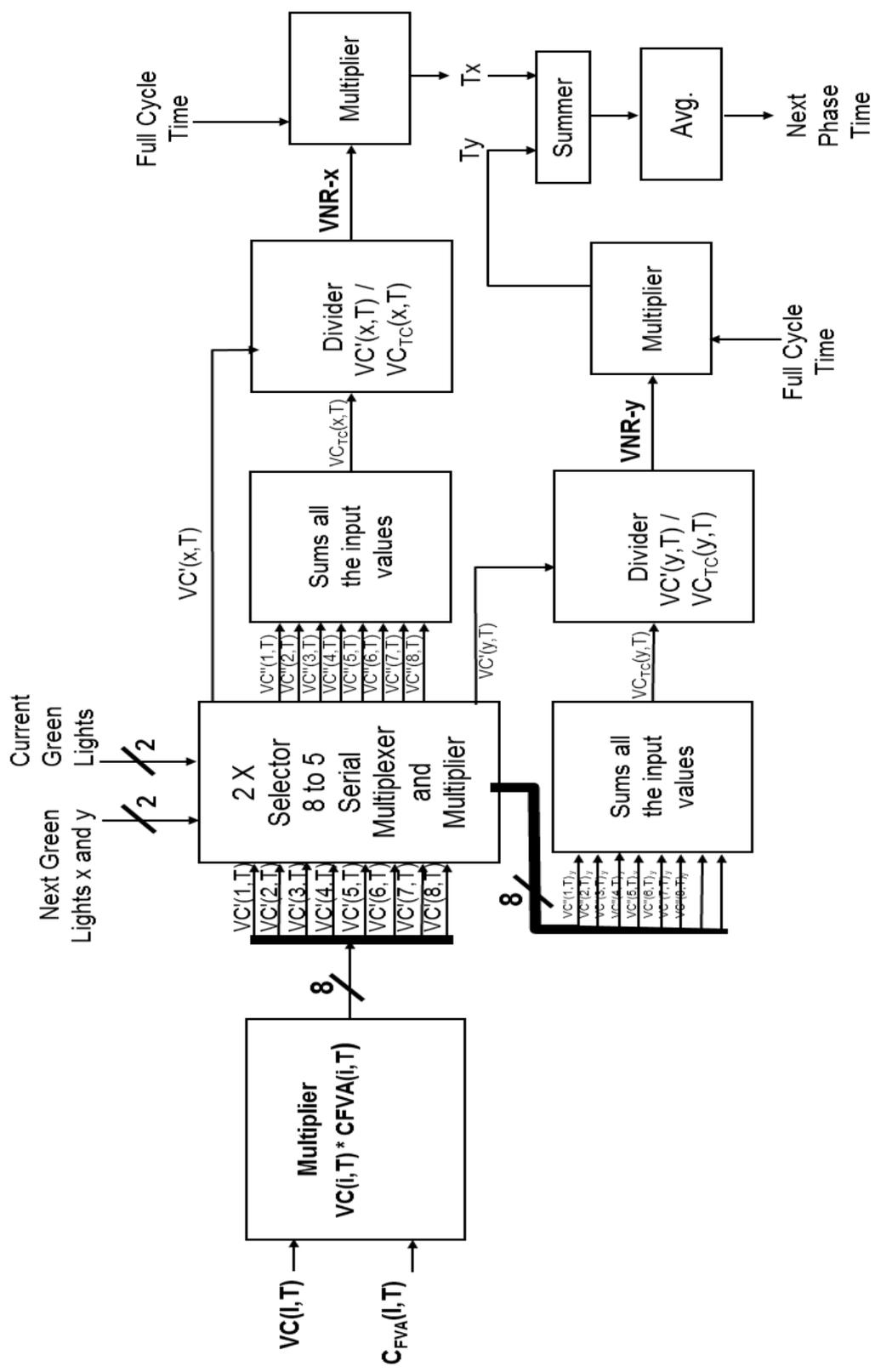

**Figure 10: The Developed Algorithm for the Next Phase Time (Internal Architecture)**

The first two blocks, in Figure 10, were added to add accuracy to the decision by choosing the only desired queue lengths to be considered in the calculations. The features of those queue lengths were: had to be waiting at the traffic light stopping line (not on the move queues), currently not green, and had to be among the competing (intersecting) queue lengths. After the calculations, illustrated in Figure 11, were made, two values of time resulted: $T_x$ (the time calculated according to the need of lane-x) and $T_y$ (the time calculated according to the need of lane-y). Since the main aim of the developed system is to distribute the time justly among the lanes, then together $T_x$ and $T_y$ were averaged to get one time value which represented the next phase time.

**4. Phase Plan Implementation**

Whenever the traffic light controller finished making the next phase plan's two decisions; the next phase green lights and the next phase time, it forwarded the phase plan to the traffic light display entities placed at the intersection to be implemented. As soon as the phase time elapsed, the traffic light controller retriggered to make a new phase plan with a new set of collected data.

**5. System Evaluation Process**

For evaluating the developed controller performance, at first, it was aimed to use the SIDRA Intersection simulator [20]. This is a standard simulator used by civil engineers; however unfortunately, it did not support the ability to customize the intersection's traffic light controller. So, the SIDRA Intersection simulator's models were adopted and built as a customized simulation tool using MATLAB [21]; then, it was validated compared with the SIDRA Intersection simulator.

During the validation of the customized simulation tool, five levels of demand were applied through five case studies on both of the simulators, SIDRA Intersection and the developed Simulation tool, using the same traffic light controller. As can be seen in Figure 12, the performances of both simulations were almost matching each other (93.3% similarity).

**Next Green Phase Time Algorithm**

1. **GET** each direction's queue length ($V_C$), the first vehicle's arrival flag ($C_{FVA}$), the two Currently Green directions' IDs ($G_{C1}$, $G_{C2}$), the selected next phase two green lights' IDs ($G_{N1}$, $G_{N2}$), and the standard Full Cycle Time.

2. **DETERMINE** which direction's queues are confirmed to have arrived:
   $V_C'(i,T) \leftarrow V_C(i,T) * C_{FVA}(i,T)$

3. **DETERMINE**, for each direction, which of its adjacent queues should be considered in calculating the next phase time whilst setting the rest (The Non-Adjacent or currently green) of them to zero.
   $V_{C(GN1)}''(i,T) \leftarrow V_C'(i,T) * $ (Is Node i adjacent to $G_{N1}$) $*$ (i~=$G_{C1}$ and i~= $G_{C2}$)
   $V_{C(GN2)}''(i,T) \leftarrow V_C'(i,T) * $ (Is Node i adjacent to $G_{N2}$) $*$ (i~=$G_{C1}$ and i~= $G_{C2}$)

4. **CALCULATE**, for each of the two directions, the summation of the results in step 3.
   $V_{CT}(G_{N1},T) \leftarrow $ Sum $(V_{C(GN1)}''(i,T))$ for i = 1 to 11
   $V_{CT}(G_{N2},T) \leftarrow $ Sum $(V_{C(GN2)}''(i,T))$ for i = 1 to 11

5. **CALCULATE**, for each of the two directions, the division of the chosen direction's queue length over the total summation found for that direction in step 4.
   $VNR_{GN1} \leftarrow V_C(G_{N1},T) / V_{CT}(G_{N1},T)$
   $VNR_{GN2} \leftarrow V_C(G_{N2},T) / V_{CT}(G_{N2},T)$

6. **CALCULATE**, for each of the two directions, what percent of the full cycle time must be given to that direction.
   $G_{TN1} \leftarrow VNR_{GN1} * $ Full_Cycle_Time (120 Seconds)
   $G_{TN2} \leftarrow VNR_{GN2} * $ Full_Cycle_Time (120 Seconds)

7. **DETERMINE** the next phase time.
   Next_Phase_Time $\leftarrow$ Average ($G_{TN1}$, $G_{TN2}$)

**Figure 11: The Developed Next Phase Time Decision Making Algorithm**

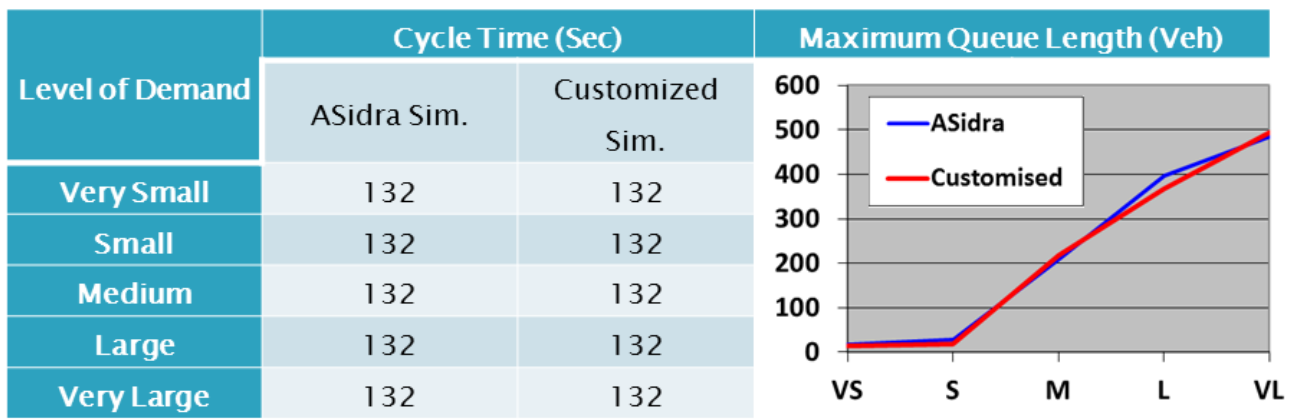

| Level of Demand | Cycle Time (Sec) | |
|---|---|---|
| | ASidra Sim. | Customized Sim. |
| Very Small | 132 | 132 |
| Small | 132 | 132 |
| Medium | 132 | 132 |
| Large | 132 | 132 |
| Very Large | 132 | 132 |

**Figure 12: Validation Results for the Customized Simulation Tool Compared to SIRDA Intersections Simulator**

After validating the developed simulation tool, it was used to evaluate the developed traffic light controller. The customized simulation tool worked as a single intersection as that shown in Figure 3.

For the evaluation purpose, four existing controllers were used to compare their performances with the developed controller. For the benchmark methods, two methods had been chosen: fixed/pre-timed phase plan control [1] and fully-actuated phase plan control [2], which were named BM1 and BM2, respectively. These two methods were chosen because they are the most used nowadays on our roads. The most related and the latest controllers to the developed controller in this paper were both of [18] and [19]. In this paper, we have referred to them as NM1 and NM2, respectively.

During the evaluation process, five levels of demand were applied to each of the five controllers (BM1, BM2, NM1, NM2, and the developed controller) running over the same four-leg intersection setup shown before. In one experiment, all of the intersection's legs received the same level of demand within an hour of the simulation time. The five levels of demand were: very-small (250 veh./hour/lane), small (375 veh./hour/lane), medium (750 veh./hour/lane), large (1125 veh./hour/lane), and very-large (1300 veh./hour/lane).

The developed controller received a set of input factors: (1) the level of demand or the arrival flow rate (pre-set), (2) queue waiting time (real-time variable), which represented the amount of time (in seconds), (3) a queue (at least one vehicle) had been waiting at red traffic light, (4) lane's occupancy % (real-time), which represented how much percentile of the whole lane was filled with vehicles, (5) $1^{st}$ vehicle's arrival flag (real-time), (6) vehicle's type (set to default), (7)on-duty flag (set to default), (8) next road occupancy (set to default), and (9) the behind road weightage (set to default). Some of the factors were set to their default values because they served special cases only.

While the outputs of the simulation tool were eight factors: (1) Departure/Arrival Ratio (Measured), (2) Max. Queue Length (Measured), (3) Avg. Queue Length (Measured), (4) Max. Queue Waiting Time (Measured), (5) Avg. Queue Waiting Time (Measured), (6) Given Green Time Utilization (Measured), (7) Stability (Observed), and (8) Overall Performance indication (Calculated). The overall performance indication factor will be calculated based on the other 7 output factors after being ranked with the help of a points-order-based ranking system. The calculation formula was the multiplication of the seventh factor (Stability) by the summation of the first six factors.

The ranking values were represented by the efficiency decremented order. Basically, it was a point-based ranking system where the methods' performances were evaluated by giving each method an amount of points that was equal to the method's performance order value.

Each experiment was repeated as many times as the sample size to assure a level of confidence of 95%. The level of confidence was mainly represented by a percentage which told how likely it was to get similar results when repeating the same experiment [22]. The experiments in this thesis were performed with a level of confidence of 95%.

Level of confidence can be calculated using equation (1).

$$n = \frac{n_0 N}{n_0+(N-1)} \quad (1)$$

Where n represents the sample size for a finite population, N is the total population, and $n_0$ is the sample size for the unknown population which can be determined from equation (2).

$$n_0 = \left(\frac{Z\sigma}{e}\right)^2 \quad (2)$$

Where the Z is the confidence level (Z=1.96 for the 95%), σ is the standard deviation which takes the default value of 0.5, and e represents the margin of error or confidence interval which is +/-5% for the 95% level of confidence (1 – 0.95 = 0.05).

According to the calculations made using equations (1) and (2), the sample sizes of 152, 190, 254, 287, and 297 samples were needed to achieve the 95% of confidence for the very-small, small, medium, large, and very large levels of demand, respectively. In other words, each experiment had to be repeated a number of times, equal to the sample size for each scenario, to be 95% confident of the results that were obtained from the valid simulation tool.

**6. Experimental Results**

All of the resulting data and the analytical results for the five experiments have been summarized and compiled into five tables (Table 1 – 5) to make them ready to be evaluated by the order-based ranking system for the overall performance evaluation.

Table 1: Experimental and Analytical Results for Experiment-1 (Very-Small Arrival Flow Rate)

| Parameters | BM1 | BM2 | NM1 | NM2 | DT3P |
|---|---|---|---|---|---|
| **Departure-Arrival Percentage** | 98.52% | 98.52% | 98.54% | 98.54% | 99.58% |
| **Avg. Q. L. (Veh)** | 6.76 | 6.76 | 2.08 | 9.71 | 2.12 |
| **Avg. W.T. (Sec)** | 85.37 | 85.33 | 18.62 | 128.54 | 17.86 |
| **Max.W.T. (Sec)** | 99.26 | 99.37 | 99.72 | 189.99 | 73.64 |
| **Max.Q. L. (Veh)** | 13.33 | 13.26 | 7.38 | 16.91 | 6.5 |
| **G. G. Time Utilization** | 0.3886516 | 0.38863115 | 0.8178481 | 0.3337493 | 0.87195 |
| **Stability (1:Yes, 0: No)** | 1 | 1 | 1 | 1 | 1 |

Table 2: Experimental and Analytical Results forExperiment-2 (Small Arrival Flow Rate)

| Parameters | BM1 | BM2 | NM1 | NM2 | DT3P |
|---|---|---|---|---|---|
| **Departure-Arrival Percentage** | 98.52% | 98.49% | 99.36% | 98.28% | 99.42% |
| **Avg. Q. L. (Veh)** | 10.11 | 10.14 | 4.39 | 14.57 | 3.87 |
| **Avg. W.T. (Sec)** | 89.94 | 89.99 | 33.68 | 133.38 | 27.14 |
| **Max.W.T. (Sec)** | 99.51 | 99.52 | 105.35 | 183.38 | 75.24 |
| **Max.Q. L. (Veh)** | 18.23 | 18.25 | 11.2 | 23.16 | 9.9 |
| **G. G. Time Utilization** | 0.530732 | 0.5313073 | 0.8257414 | 0.4693992 | 0.938666 |
| **Stability (1:Yes, 0: No)** | 1 | 1 | 1 | 1 | 1 |

Table 3: Experimental and Analytical Results for Experiment-3 (Medium Arrival Flow Rate)

| Parameters | BM1 | BM2 | NM1 | NM2 | DT3P |
|---|---|---|---|---|---|
| **Departure-Arrival Percentage** | 70.16% | 70.18% | 92.13% | 97.20% | 97.26% |
| **Avg. Q. L. (Veh)** | 121.21 | 120.9 | 39.18 | 30.49 | 15.74 |
| **Avg. W.T. (Sec)** | 98.81 | 98.8 | 68.73 | 139.89 | 82.43 |
| **Max.W.T. (Sec)** | 100.05 | 100.13 | 263.01 | 226.44 | 134.02 |
| **Max.Q. L. (Veh)** | 215.4 | 214.99 | 135.56 | 32.84 | 24.97 |
| **G. G. Time Utilization** | 0.989945 | 0.9899303 | 0.9890814 | 0.91325509 | 0.997515 |
| **Stability (1:Yes, 0: No)** | 1 | 1 | 1 | 1 | 1 |

Table 4: Experimental and Analytical Results for the Experiment-4 (Large Arrival Flow Rate)

| Parameters | BM1 | BM2 | NM1 | NM2 | DT3P |
|---|---|---|---|---|---|
| **Departure-Arrival Percentage** | 67.32% | 67.14% | 61.88% | 69.26% | 68.76% |
| **Avg. Q. L. (Veh)** | 196.21 | 198.73 | **788.85** | **798.75** | 185.49 |
| **Avg. W.T. (Sec)** | 98.9 | 98.89 | **2386.54** | **2511.09** | 107.25 |
| **Max.W.T. (Sec)** | 100.45 | 100.34 | **3182.06** | **3348.13** | 144.98 |
| **Max.Q. L. (Veh)** | 347.5 | 355.6 | **1051.8** | **1065.38** | 331.36 |
| **G. G. Time Utilization** | 0.989804 | 0.9897790 | 0.9949038 | 0.92669651 | 0.998982 |
| **Stability (1:Yes, 0: No)** | 1 | 1 | 0 | 0 | 1 |

Table 5: Experimental and Analytical Results for Experiment-5 (Very-Large Arrival Flow Rate)

| Parameters | BM1 | BM2 | NM1 | NM2 | DT3P |
|---|---|---|---|---|---|
| **Departure-Arrival Percentage** | 58.34% | 58.34% | 53.59% | 60.50% | 59.32% |
| **Avg. Q. L. (Veh)** | 282.45 | 282.85 | **932.955** | **941.265** | 273.37 |
| **Avg. W.T. (Sec)** | 98.93 | 98.93 | **2598** | **2660.4** | 105.16 |
| **Max.W.T. (Sec)** | 100.56 | 100.93 | **3464** | **3547.2** | 138.36 |
| **Max.Q. L. (Veh)** | 518.68 | 525.51 | **1243.94** | **1255.02** | 511.31 |
| **G. G. Time Utilization** | 0.991463 | 0.9914038 | 0.9955182 | 0.92669651 | 0.999137 |
| **Stability (1:Yes, 0: No)** | 1 | 1 | 0 | 0 | 1 |

Starting from the Very-Small arrival rate, Table 1, till the Medium arrival flow rate, Table 3, it can be seen that both the NM1 and DT3P achieved better results than the benchmark methods in maintaining the Maximum Queue Length and the Maximum Waiting Time. DT3P continued to perform well through the fourth and fifth experiments, Table 4 and Table 5; whilst, both the NM1 and NM2 methods totally lost their stability, this is why their performance at the Large and Very-Large demand levels had to be neglected. From Table 1 through Table 5, it can be seen that DT3P reduced the Maximum and the Average Queue Length better than all of the other methods at all of the levels. It peaked at the medium level of demand, with the least effect on the Maximum waiting time, unlike NM1 and NM2, which lost their control at the large and the very-large arrival flow rates.

The overall performances of all the five controllers can be seen in Figure 13. It is observable that both of NM1 and NM2 lost their stability at the Large and Very-Large levels of demand. Unlike NM2, NM1, which performed better than BM1 and BM2 at the first two levels of demand. The best recorded performance for NM2 can be seen at the medium level of demand. DT3P peaked at the full score of 30 points at the small level of demand. At all levels of demand, DT3P performed better than the other controlling methods.

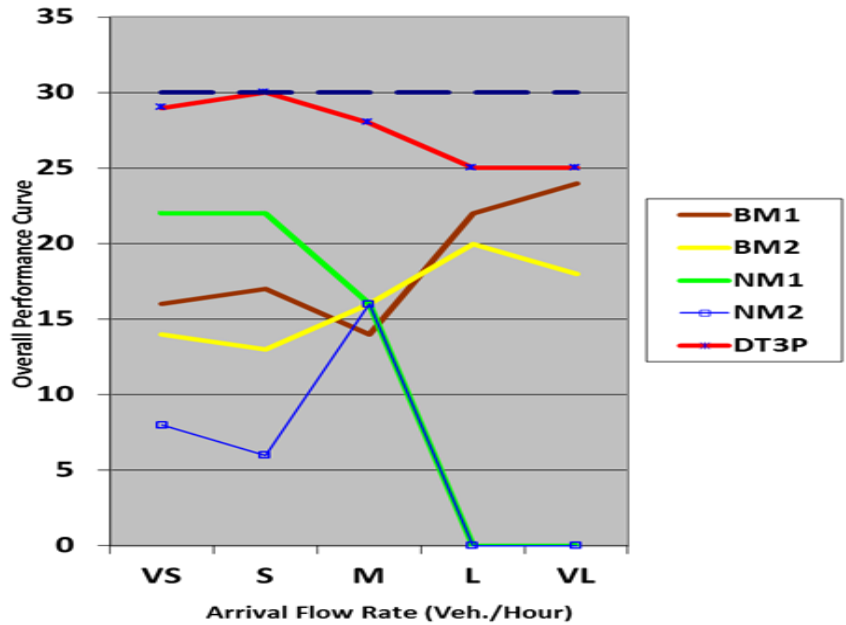

**Figure 13: Overall performance for all the five controlling methods**

**7. Conclusions**

In this paper a traffic light controlling method was introduced and its algorithms were illustrated in details. It was introduced as the developed Dynamic Traffic Light Phase Plan Protocol (DT3P), which was then tested and compared with four other controlling methods (BM1, BM2, NM1, and NM2). The first two methods are the benchmark methods which we can find in every city even though they have been introduced in 1970s. While both of NM1 and NM2 are new methods those were introduced in 2010. The test was done using a customized simulation tool developed by the research team using MATLAB programing and was validated using SIDRA Intersections simulator. A set of 25 experiments were held (five levels of demand were applied into each of the five controllers). Each experiment was replicated according to the calculated number of runs to achieve a level of confidence of 95%. The comparison showed the efficiency of the developed method which can be seen through its anticipation in increasing the intersection throughput when it increased the green time utilization, decreasing the avg. and max. Queue lengths, and its anticipation in optimizing the avg. and max. Queue's waiting time according to the needs. Another finding was that both NM1 and NM2 were ineligible to be used at intersections with large and very large levels of demand. Finally, DT3P was found eligible to be used at any intersection with all levels of demand.

**Acknowledgement**

This research was financially supported by Universiti Teknologi Petronas.

**Maythem Kamal Abbas** was born in Baghdad, Iraq, in 1984. He got his BSc in Control and Computer Systems Engineering from University of Technology in Baghdad, Iraq in 2005. Subsequently he completed his Higher Diploma in IT in Bickenhall College of Computing (Middlesex University partner) in London, United Kingdom in 2008. Consequently, he completed his Master's degree in IT from Universiti Teknologi Petronas, Malaysia in 2009. Currently, he is a doctor of philosophy (PhD in Electrical and Electronic Engineering) in Universiti Teknologi Petronas. Since 2008, he is a researcher with Information Technology Department and Electrical and Electronics Engineering department in Universitit teknologi Petronas. He is the author of 12 conference and Journal publications. His research interests include vehicular ad hoc networks, formal specification language, decision making algorithms, protocol design, intelligent systems and robotics. Since 2005, he is an active member of the Iraqi Engineers Union (IEU). He served as a reviewer and as a technical programme committee in many international conferences and journals.

**Mohd Noh Karsiti** completed his training and obtained the degrees of Bachelor of Science in Electrical Engineering and Masters of Science in Electrical Engineering from California State University, Long Beach, USA, in 1985 and 1987 respectively. Subsequently, he completed his doctoral program in University of California, Irvine and was awarded PhD in Electrical and Computer Engineering in 1991.

**Madzlan Napiah** is a lecturer in the Civil Engineering Department Universiti Teknologi PETRONAS. He holds a PhD in Civil Engineering and MSc in Transportation Planning & Engineering, both from the University of Leeds UK. He also holds a BSc in Civil Engineering from Michigan State University USA. His research interests are in modelling and optimisation of public transportation, traffic flow, and also in highway materials and project planning and management.

**Samir Brahim-Belhaouari** received the Master's degree in Telecommunications in 2000 from Institut Nationale Polytechnique (INP) of Toulouse, France and the PhD in 2006 from the Federal Polytechnic School of Lausanne (EPFL). He has been Senior Lecturer at Universiti Teknologi PETRONAS since 2007. His research interests lie in the areas of image and speech processing and stochastic processes.

**Marwan Al-Jemeli** Received his BSc (2005) in computer engineering from University of Baghdad/college of Engineering, Baghdad, Iraq. He received his MSc (2010) and PhD (2014) in Electrical and electronic engineering from Universiti Teknologi PETRONAS, Perak, Malaysia. Since 2010 he is a research scholar at The Centre for Intelligent Signal and Imaging Research (CISIR). His current research interests includes: embedded systems, computer networks, wireless network, wireless sensor networks, networks routing, medium access control, localization and location estimation and mobility in wireless networks.